\documentclass[11pt,a4paper]{article}
\usepackage{float}     
\usepackage{placeins}  

\usepackage{xspace}
\usepackage{comment}
\usepackage{subcaption}
\usepackage{graphicx}
\usepackage{tikz}
\usetikzlibrary{arrows.meta, positioning, shapes.multipart, calc}

\title{Unfolding the Energy Spectrum of Ultra-High-Energy Cosmic Rays Using Pierre Auger Open Data}

\author{
Ji\v{r}\'{i} Kvita$^{1}$ \and
Petr Baro\v{n}$^{1}$
}

\date{
$^{1}$Palacky University Olomouc, Faculty of Science, Joint Laboratory of Optics of Palacky University and Institute of Physics of the Czech Academy of Sciences,\\
17. listopadu 1192/12, 779 00 Olomouc, Czech Republic\\[1ex]
\texttt{petr.baron@upol.cz, jiri.kvita@upol.cz}
}

\begin{document}
\maketitle

\begin{abstract}
We reconstruct the energy spectrum of ultra-high-energy cosmic rays using the publicly released Pierre Auger Observatory data set. 
Since event-level Monte Carlo truth information is not included in the open data, we develop a consistent procedure to regenerate a pseudo-Monte Carlo sample directly from the published quantities: the registered event counts $N$, the unfolded spectrum $N_\textrm{corr}$, and the detector response matrix $R_{ij}$ from the Auger 2020 spectrum data analysis.
Using the row-normalized response matrix and the published unfolded spectrum as a truth prior, we construct an absolute-level migration matrix and generate the event-by-event truth and reconstructed-level pairs by drawing from a two-dimensional probability distribution function. The resulting sample statistically replicates the detector response properties of the Pierre Auger Surface Detector.
This pseudo-MC sample allows for the application of classical unfolding techniques (bin-by-bin and iterative Bayesian unfolding via \texttt{RooUnfold}) as well as a machine-learning-based unfolding using \texttt{OmniFold}. 
We demonstrate that using such publicly available information this approach allows the full unfolding procedure.
\end{abstract}





\section{Introduction}

Ultra-high-energy cosmic rays (UHECRs) provide a unique probe of extreme astrophysical environments, raising questions of the very origin of UHECR, source, chemical composition, but also testing our understanding of their propagation and interaction with the atmosphere.

The Pierre Auger Observatory~\cite{PierreAuger:2015eyc} has measured the cosmic-ray flux above $10^{18}\,$eV with 
unprecedented precision~\cite{PierreAuger:2020qqz} and presents the results as energy spectrum corrected for the effects of finite detector resolution, a procedure known as unfolding.
Its open-data release includes reconstructed event information yet not the Monte Carlo truth spectrum used in their unfolding procedure.
The unfolding as performed in the Auger spectrum analysis -- relying 
on full detector simulations -- cannot therefore be exactly reproduced 
directly using only the open data.

Motivated by checking the usefulness of the public data provided by Auger, we set on attempting the unfolding procedure using the published information, develop a method of performing both the traditional as well as machine learning (ML) based unfolding and compare to the published results.

\section{Methodology}

In order to overcome the direct missing information at the generator level, we design the following procedure.
The 2020 Auger energy-spectrum paper~\cite{PierreAuger:2020qqz} provides three essential public ingredients necessary for spectral unfolding:
\begin{enumerate}
    \item the measured counts $N_i$ in each energy bin;
    \item the unfolded (acceptance-corrected) counts $N_{\textrm{corr},i}$ in each energy bin;
    \item the detector response matrix $R_{ij}$ describing the bin-to-bin migrations due to finite experimental energy resolution.
\end{enumerate}

\begin{figure}[p]
\centering
\begin{tikzpicture}[
    node distance=1.4cm and 2.2cm,
    box/.style={
        rectangle,
        draw=black,
        fill=blue!10,
        rounded corners,
        align=center,
        minimum width=4.4cm,
        minimum height=1.0cm,
        font=\small
    },
    arrow/.style={
        -{Stealth[scale=1.2]}, thick
    }
]

\node[box] (N)       {{\bf Measured counts}\\$N_i$};
\node[box, right=of N] (Ncorr) {{\bf Unfolded counts}\\$N_{\textrm{ corr},i}$};
\node[box, below=1.7cm of $(N)!0.5!(Ncorr)$] (R)
        {{\bf Published response matrix}\\$R_{ij}$ (row-normalised)};
        
\node[box, below=1.7cm of R] (M)
        {{\bf Absolute migration matrix}\\$M_{ij} = R_{ij}\,N_{\textrm{ corr},j}$};

\node[box, below=1.7cm of M] (MC)
        {{\bf Pseudo-MC 2D generation}\\ of (truth, reco) pairs};

\node[box, below left=1.7cm and 0.5cm of MC] (Bayes)
        {{\tt\bf RooUnfold}\\Bayes};

\node[box, below=1.7cm of MC] (BinByBin)
        {{\tt\bf RooUnfold}\\Bin-by-bin};

\node[box, below right=1.7cm and 0.5cm of MC] (Omni)
        {{\tt\bf OmniFold}\\ML unfolding};

\node[box, below=2.0cm of BinByBin] (Final)
        {{\bf Unfolded spectrum}\\compared to $N_{\textrm{corr}}$};

\draw[arrow] (N) -- (R);
\draw[arrow] (Ncorr) -- (R);
\draw[arrow] (R) -- (M);
\draw[arrow] (M) -- (MC);

\draw[arrow] (MC) -- (Bayes);
\draw[arrow] (MC) -- (BinByBin);
\draw[arrow] (MC) -- (Omni);

\draw[arrow] (Bayes) -- (Final);
\draw[arrow] (BinByBin) -- (Final);
\draw[arrow] (Omni) -- (Final);

\end{tikzpicture}
\caption{Flow diagram of the preparatory work and unfolding methods used in this work. Measured counts $N_i$, unfolded reference counts $N_{\mathrm{corr},i}$, and the published response matrix $R_{ij}$ are combined to construct an absolute migration matrix $M_{ij}$. 
A pseudo-MC sample is generated by randomly sampling a two-dimensional histogram, enabling both classical and machine-learning unfolding approaches, results of which are then compared to the published unfolded spectrum.}
\label{fig:flow}
\end{figure}
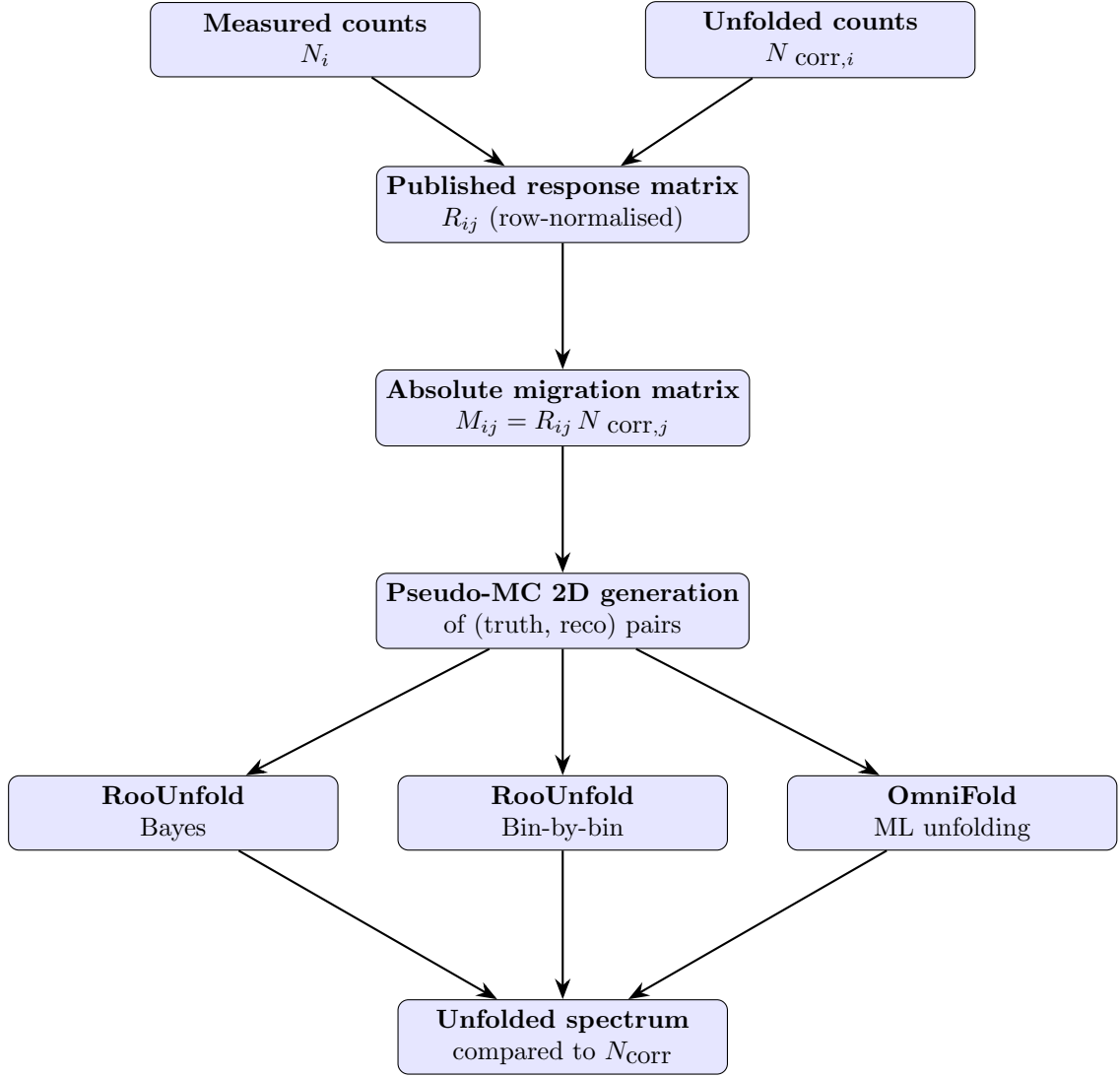

Still, the fully populated (absolute) migration matrix is missing and so there is no direct access to the generator-level spectrum. Still, we argue that the presented quantities allow one to design a model to reconstruct a pseudo-Monte Carlo sample statistically equivalent to the original simulation used by Auger, making it possible to perform unfolding techniques such as \texttt{OmniFold}~\cite{Andreassen:2019cjw}.
Namely, in this work we perform the following steps:
\begin{itemize}
    \item construct the detector migration response matrix $R_{ij}$ from the published table;
    \item build the absolute migration matrix $M_{ij}=R_{ij}N_{\mathrm{corr},j}$ using the published unfolded spectrum $N_{\mathrm{corr},j}$;
    \item generate a two-dimensional pseudo-MC correlated truth--reco pairs (pseudo-events) by randomly sampling the migration matrix;
    \item unfold the open-data measured spectrum using \texttt{RooUnfold} and both the published as well as the generated pseudo-data using the \texttt{OmniFold} technique;
    \item compare the unfolding results obtained using the different methods to the published $N_\textrm{ corr}$.
\end{itemize}
The diagram depicted in Figure~\ref{fig:flow} summarizes the unfolding workflow used in this analysis. In the following, we describe the steps in more detail.

\section{Reconstruction of the Migration Matrix}

The Auger response matrix $R_{ij}$ describes the probability that an event originating in the true energy bin $j$ is reconstructed in detector-level bin $i$. 
Since the open data do not provide the truth-level MC events, we reconstruct the absolute migration matrix as
\begin{equation}
M_{ij} = R_{ij}\,N_{\mathrm{corr},j},
\end{equation}
which represents the expected number of events migrating between true and reconstructed bins according to the published information.

This matrix is then interpreted as a two-dimensional probability distribution. 
Using the \verb|TH2::GetRandom2| method as provided by the ROOT framework~\cite{ROOT_NIMA_1997}, we generate Monte Carlo events in terms of  correlated truth--reco pairs. 
We create a sample containing ten times as many events as the open-data set to ensure sufficient statistical power for the \texttt{OmniFold} training.

\section{Unfolding Methods}

In order to evaluate the performance of the ML-based unfolding method, we compare it to two classical and widely used methods. We thus utilize the three unfolding techniques as listed below.

\paragraph{Bin-by-bin:}  
A direct correction using the ratio of truth to reconstructed MC yields a quick and robust method but migration-unaware in the strict sense and sensitive to the requirement of the MC spectrum be close to the one in data.

\paragraph{Bayesian unfolding:}  
We use the D'Agostini's iterative Bayesian method as implemented in \texttt{RooUnfold}~\cite{Adye:2011gm}. 
With four iterations, this method accurately reproduces the Auger published unfolded spectrum.

\paragraph{OmniFold:}  
A machine-learning based reweighting method using iterative classifiers to match both detector-level and truth-level distributions. 

The presented pseudo-MC sample provides the required truth--reco pairs for the training phase to the algorithm.
Both the detector-level and generator-level classifiers used in the OmniFold procedure have identical fully connected neural-network architectures, implemented using the \texttt{energyflow.archs.DNN} backend.

Each network takes as input a single standardized scalar feature corresponding to the reconstructed or true logarithmic energy, $\log_{10}(E/\mathrm{eV})$.
The input layer therefore has dimensionality one.
This is followed by two hidden dense layers, each consisting of 100 neurons with ReLU activation functions~\cite{relu}, providing sufficient representational capacity to model non-linear decision boundaries while avoiding over-parameterization.

The output layer consists of two neurons with a softmax activation, yielding class probabilities
for binary classification (data versus Monte Carlo at detector level, and reweighted versus
original Monte Carlo at generator level).

The total number of trainable parameters per network is 10502, dominated by the fully connected weight matrices between the hidden layers.
All parameters are trainable, with no frozen layers or explicit regularization terms applied.
The networks are optimized using the categorical cross-entropy loss function and the Adam optimizer~\cite{adam} with default EnergyFlow settings.

This lightweight architecture was chosen to ensure training stability and to reduce susceptibility to statistical fluctuations, given the one-dimensional nature of the unfolding observable and the limited size of the available open-data sample.

\section{Results}

\begin{figure}[p]
  \centering
  \includegraphics[width=0.7\textwidth]{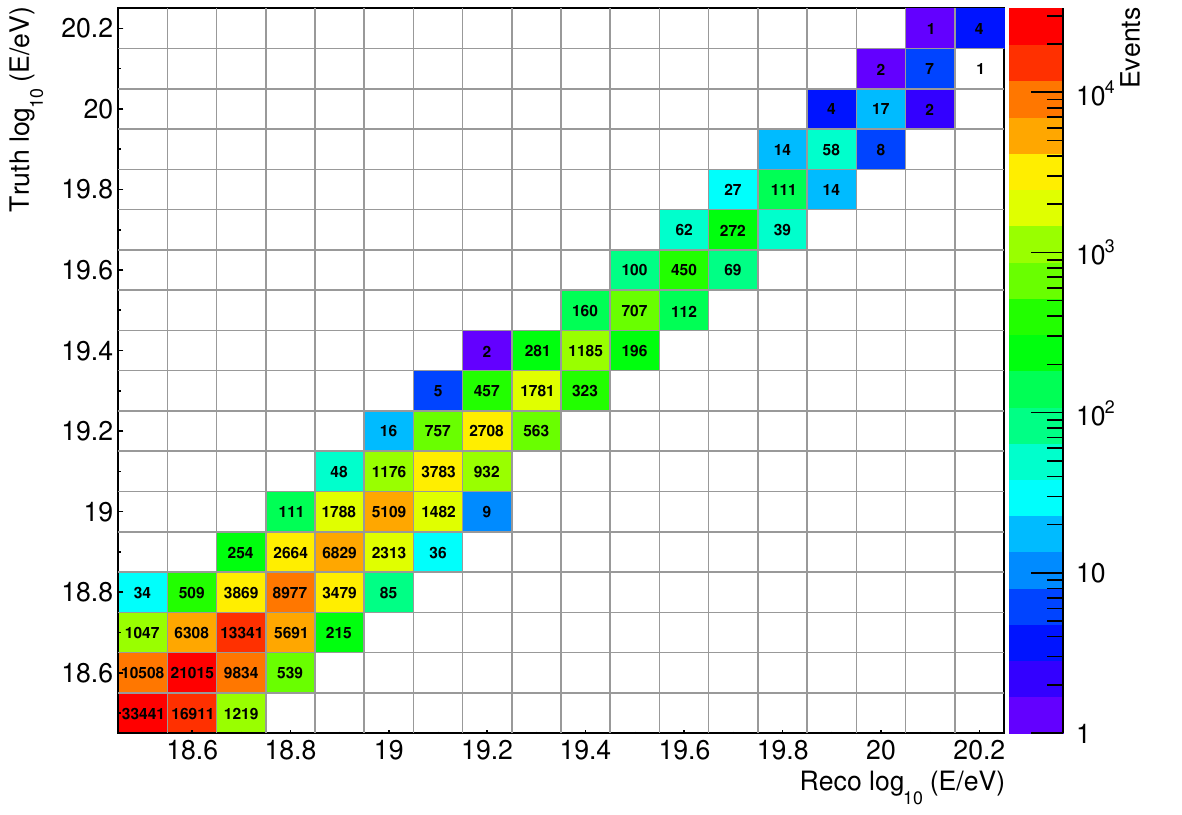}
  \caption{
Reconstructed absolute migration matrix 
$M_{ij} = R_{ij}\,N_{\mathrm{corr},j}$ obtained by scaling the published Pierre Auger row-normalized response matrix $R_{ij}$ with the unfolded spectrum $N_\mathrm{corr}$ from Table~VI of Ref.~\cite{PierreAuger:2020qqz}. 
This matrix represents the expected number of events migrating between true and reconstructed energy bins and forms the basis for generating a self-consistent pseudo-Monte Carlo dataset.
}
  \label{fig:migration_original}
\end{figure}

\begin{figure}[p]
  \centering
  \includegraphics[width=0.7\textwidth]{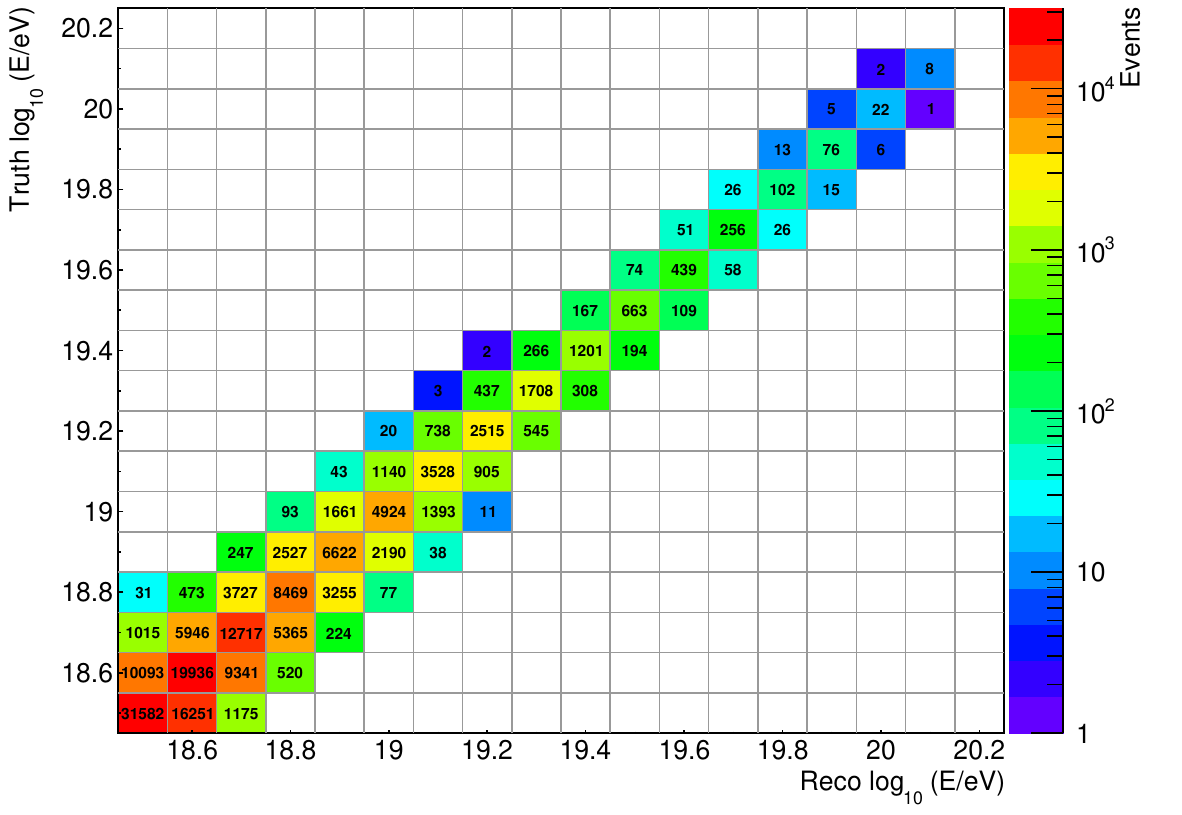}
  \caption{
Two-dimensional distribution of the (truth,reconstructed) 
energies for pseudo--Monte Carlo events generated from the original migration matrix $M_{ij}$ by randomly sampling a two-dimensional histogram. 
This sampling converts the expected migration rates into an event-level dataset with correlated truth and reconstructed energies. 
This pseudo--MC sample is used to train the \texttt{OmniFold} unfolding as well as to validate classical unfolding methods.
}
  \label{fig:migration_pseudomc}
\end{figure}

\begin{figure}[p]
  \centering
  \includegraphics[width=0.65\textwidth]{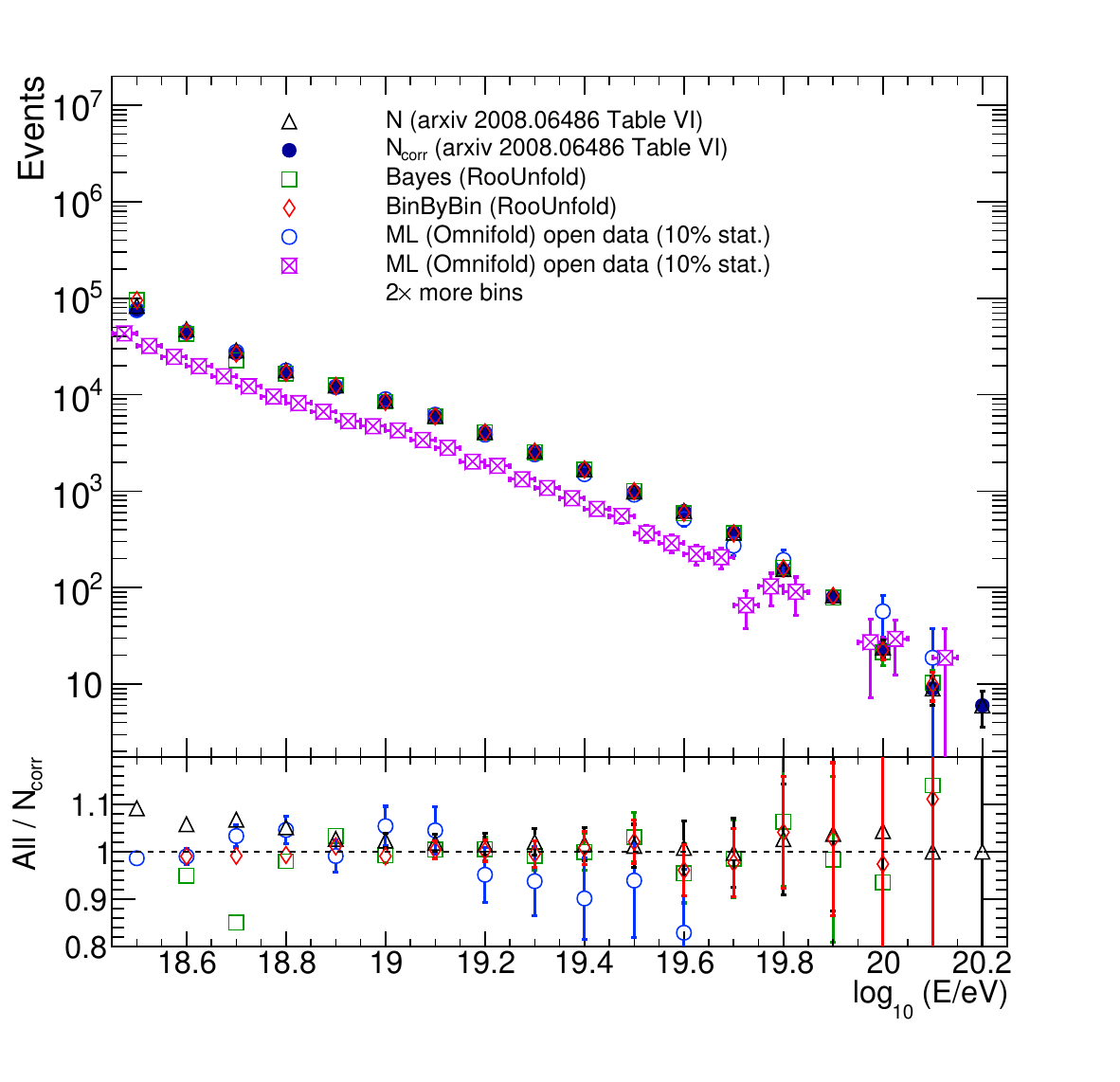}
  \caption{
Comparison of the EAS unfolded energy spectra obtained using several methods. 
Shown are the measured counts $N$, the published unfolded spectrum $N_\mathrm{corr}$, the \texttt{RooUnfold} Bayesian and bin-by-bin results using the 
reconstructed response matrix, and the \texttt{OmniFold} machine-learning unfolding of the 10\% Pierre Auger Open Data sample. 
A second, doubly-binned, \texttt{OmniFold} curve is included to emphasize the possibility of finer binning choice. 
The lower panel displays the ratio of all unfolded spectra to 
$N_\mathrm{corr}$.
}
  \label{fig:unfold_counts}
\end{figure}

\begin{figure}[p]
  \centering
  \includegraphics[width=0.65\textwidth]{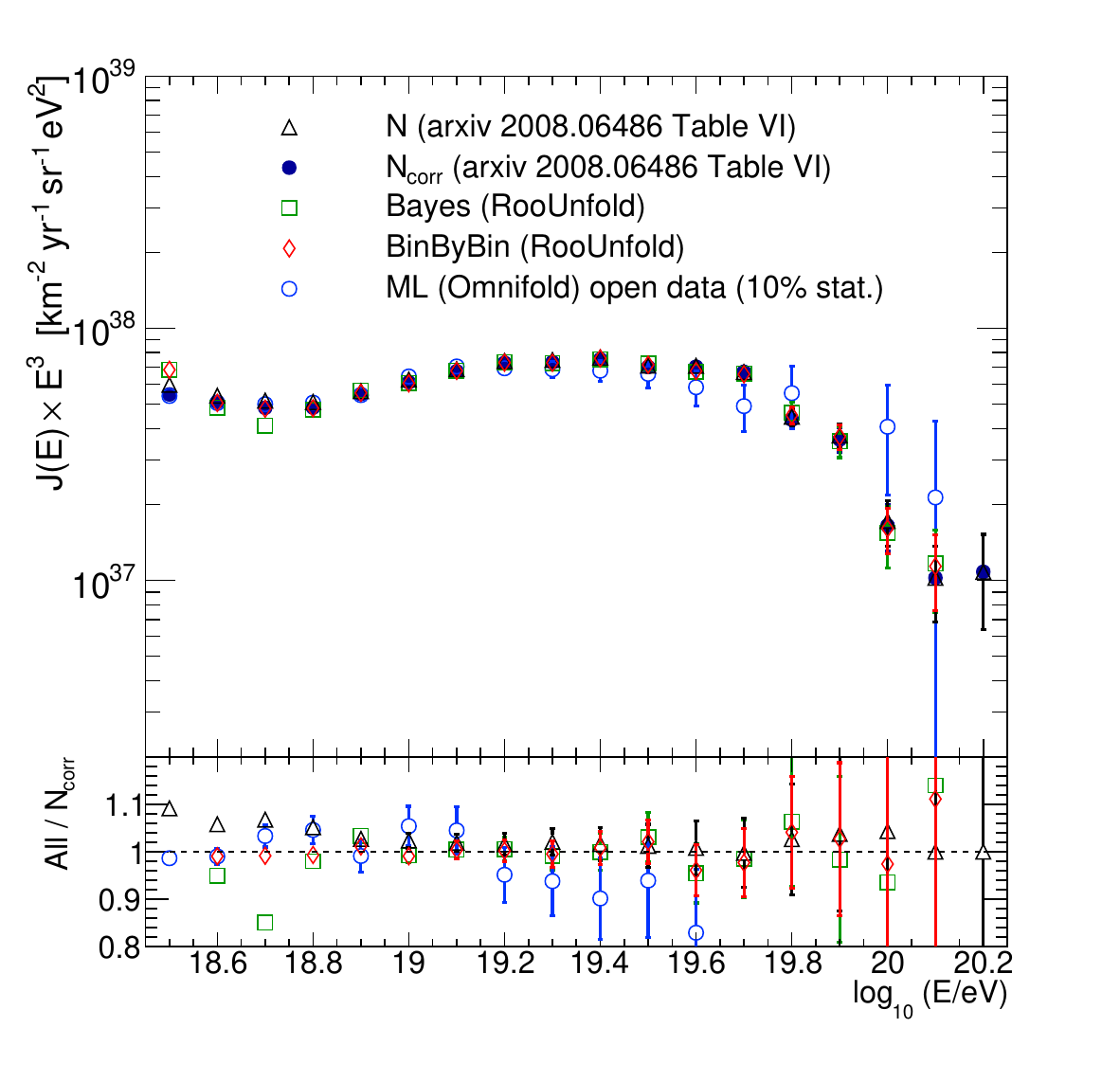}
  \caption{
The binned EAS energy spectrum expressed as $J(E)\,E^{3}$ in the measured counts $N$, 
the published unfolded spectrum $N_\mathrm{corr}$, the \texttt{RooUnfold} Bayesian and bin-by-bin results, and the \texttt{OmniFold} unfolding of the 10\% open-data sample. 
Multiplication by $E^{3}$ reduces the dynamic range of the flux and highlights differences in the spectral shape. 
The lower panel shows the ratio to $N_\mathrm{corr}$.
}
  \label{fig:unfold_flux}
\end{figure}

The reconstructed migration matrix, the corresponding pseudo-Monte Carlo sample, and the unfolded spectra obtained with different methods are summarized in Figures~\ref{fig:migration_original}--\ref{fig:unfold_flux}.

Figure~\ref{fig:migration_original} shows the absolute migration matrix $M_{ij}$ obtained by scaling the published response matrix with the unfolded spectrum $N_\mathrm{corr}$. 
The matrix populated in this way reflects the causal flow due to the finite detector resolution inducing event migrations from each true to another reconstructed energy bin. 
Figure~\ref{fig:migration_pseudomc} displays a similar distribution but now obtained from the generated pseudo-event, \emph{i.e.} by randomly sampling the two-dimensional original histogram, showing a good agreement to the original one.

The pseudo-MC provides the (truth,reconstruction)-level pairs with realistic migration patterns. Generating ten times the statistical power of the open-data sample, a stable \texttt{OmniFold} training and a robust validation procedure is possible and can be compared to classical unfolding approaches.

Figure~\ref{fig:unfold_counts} compares the unfolded event-count spectra obtained using bin-by-bin unfolding, iterative Bayesian unfolding, and \texttt{OmniFold}.

First, one notices the ratio of the detector-level spectrum to the unfolded one (black points) indicates that a correction of about 10\% is needed at lower energy.  
It can be seen that both the Bin-by-bin as well as the Bayes methods exhibit a non-closure of up to 20\% in first three bins.
The ML-based unfolding describes the lower energy spectrum part better than the two classical methods, although it later arrives to higher fluctuations and a bias of 5--10\% at higher energies, but flat. Still, the doubly-binned ML-unfolded spectrum remains smooth. The difficulty for the ML technique is the fact that it is trained only on provided open data which constitute 10\% of the full data  samples as in~\cite{PierreAuger:2020qqz}.

All methods reproduce the reference spectrum $N_\mathrm{corr}$ from Table~VI of Ref.~\cite{PierreAuger:2020qqz} within uncertainties. 
The agreement confirms that the reconstructed migration matrix and pseudo-MC generation are both accurate and self-consistent for the studied purposes. 

The \texttt{OmniFold} result, while based only on the 10\% open-data subset, follows the classical unfolding results reasonably well, although it starts to deviate from the published spectrum at higher energies as the steeply falling spectrum quickly runs out of events. Still, this is an explicit demonstration that the machine-learning approach can operate reliably even in low-statistics conditions.

Figure~\ref{fig:unfold_flux} presents the unfolded flux in the form $J(E)\,E^{3}$, which compresses the dynamic range and reveals the spectral shape more clearly. 

\begin{figure}[p]
  \centering
  \includegraphics[width=0.7\textwidth]{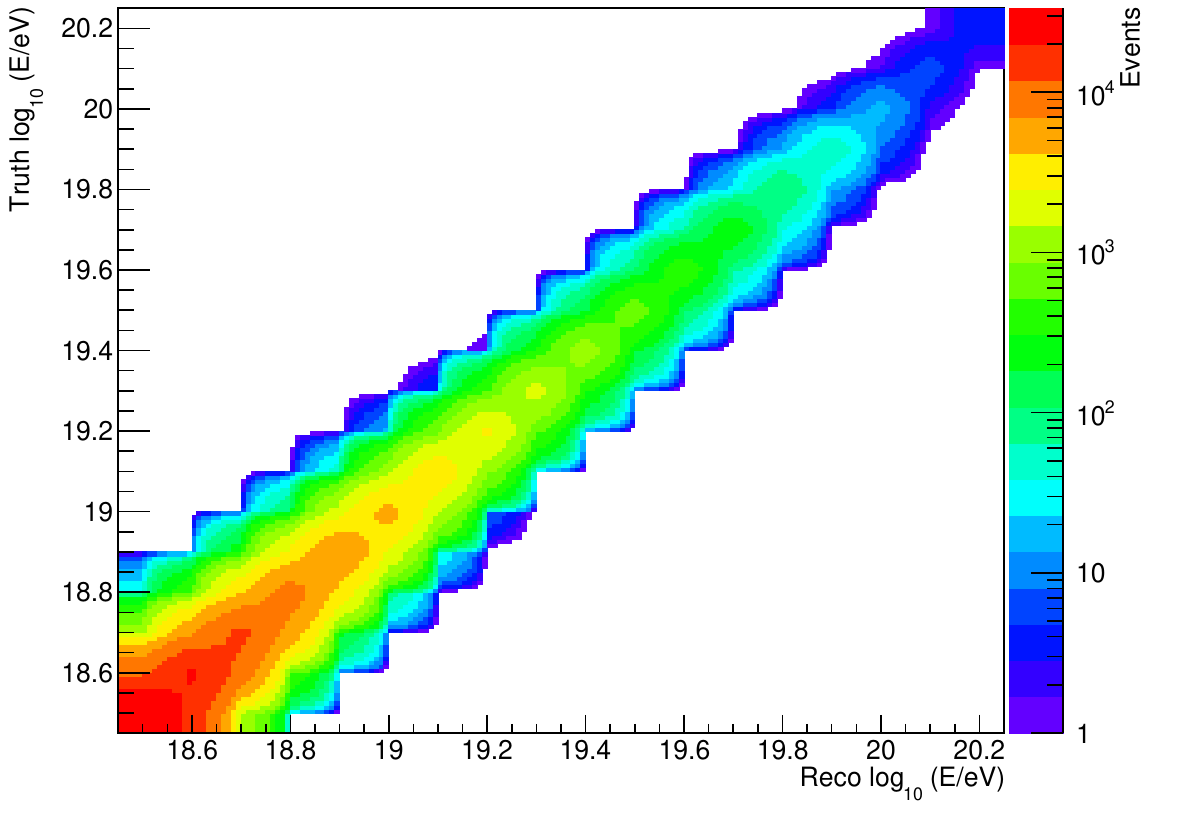}
  \caption{
The smoothened version of the two-dimensional distribution of the (truth,reconstructed) energies for pseudo--Monte Carlo events generated from the original migration matrix $M_{ij}$ by smoothly randomly sampling a two-dimensional histogram. 
This sampling converts the expected migration rates into an event-level dataset with correlated truth and reconstructed energies. 
This pseudo--MC sample is used to train the \texttt{OmniFold} unfolding as well as to validate classical unfolding methods.
}
  \label{fig:migration_pseudomc_smooth}
\end{figure}

\begin{figure}[p]
  \centering
  \includegraphics[width=0.7\textwidth]{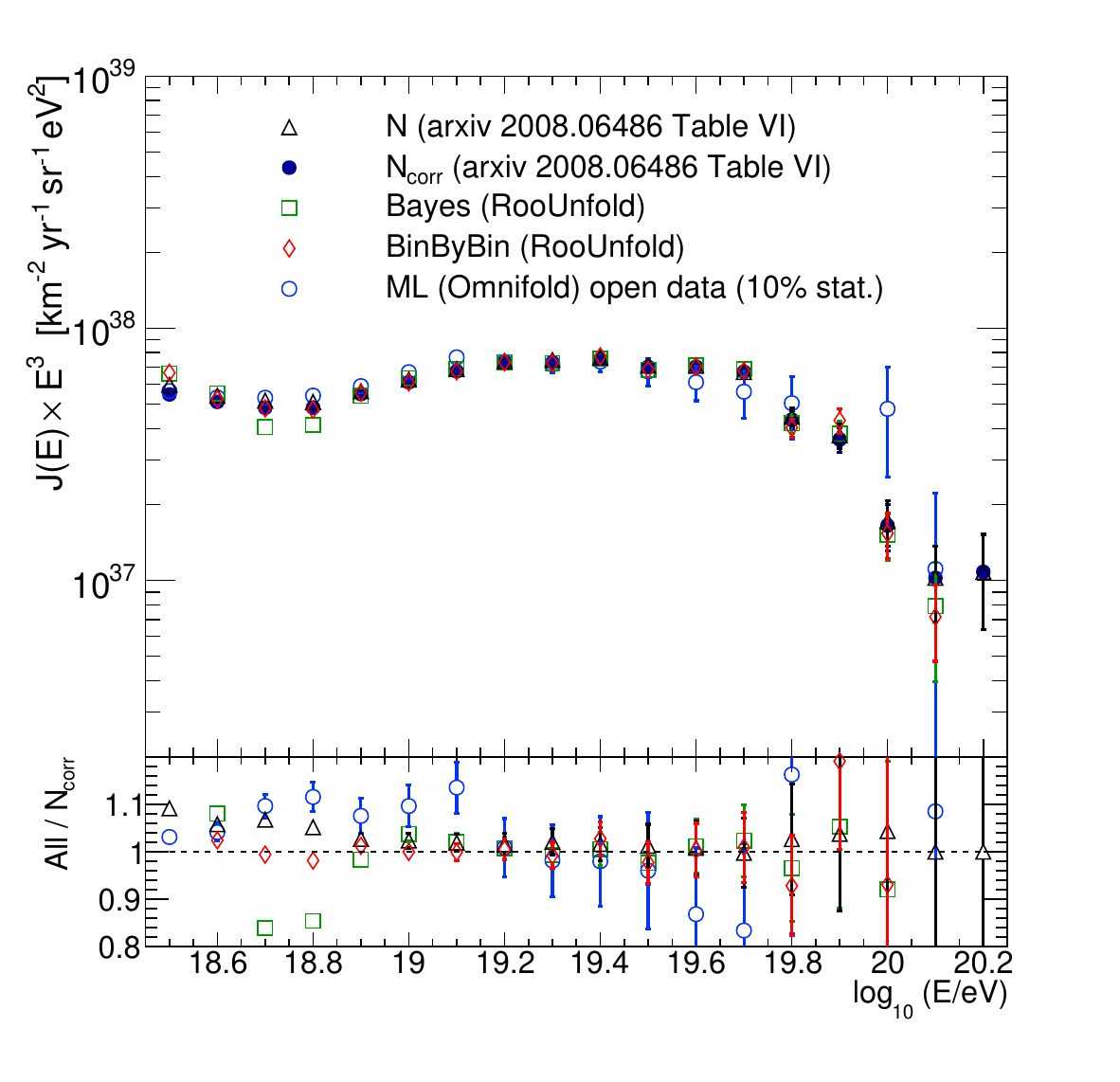}
  \caption{
Now with the ML result using a smoothened migration matrix, the binned EAS energy spectrum expressed as $J(E)\,E^{3}$ in the measured counts $N$, 
the published unfolded spectrum $N_\mathrm{corr}$, the \texttt{RooUnfold} Bayesian and bin-by-bin results, and the smooth version of the \texttt{OmniFold} unfolding of the 10\% open-data sample. 
The lower panel shows the ratio to $N_\mathrm{corr}$.
}
  \label{fig:unfold_flux_smooth}
\end{figure}

The limitation of the GetRandom2 function is in providing a distribution which is not smooth.
In order to check whether the OmniFold performance improves upon training using more smooth event, the ML exercise was repeated with such a smoothing. The results are presented in~Figure~\ref{fig:migration_pseudomc_smooth} as for the populated smoother migration matrix ans finally as the scaled flux as in~Figure~\ref{fig:unfold_flux_smooth}, showing a worse level of agreement compared to the original OmniFold training method. This suggests either the method instability, \emph{i.e.} that a larger data set is needed to be reliably unfolded, or that a lager realistic set is needed to generate the training sample in the first place.


\section{Conclusions}

All unfolding methods yield compatible spectral features across the full energy range. 
The excellent agreement in both the event-count and flux 
representations confirms the internal consistency of the unfolding chain and validates the use of pseudo--MC samples for ML-based unfolding in the absence of full detector simulations.

Using only the publicly available Pierre Auger information, we successfully reproduced the UHECR unfolded spectrum through classical and machine-learning methods. We thus demonstrate that the published response matrix, combined with open data, allows for a complete detector-model reconstruction.  

All three unfolding techniques used in this study yield unfolded spectra in good agreement with the published Auger results.
\texttt{OmniFold} unfolding applied to UHECR data is shown to be of reasonable potential. It seems to be limited by the small publicly available sample (10\% of the Auger 2020 data) and possibly also the training set from which the pseudo MC pairs of truth and reconstructed level are drawn. It still might offer a promising tool for future analyses using larger open data sets.

These results demonstrate that the methodology developed in this work successfully reproduces the published Pierre Auger energy spectrum and provides a consistent framework for applying machine-learning-based unfolding to the open data set.

\section{Acknowledgments}
The authors would like to thank the Czech Science Foundation project GA\v{C}R 23-07110S for the support of this work. We thank P.~Trávníček and V.~Novotný for useful discussions.

\bibliography{main}{}
\bibliographystyle{unsrt}


\end{document}